\documentclass[a4paper,11pt]{article}
\pdfoutput=1 

\usepackage{jcappub} 

\usepackage[T1]{fontenc} 
\usepackage{amstext,amssymb}
\usepackage{amsmath}
\usepackage{graphicx}
\usepackage{xspace}
\usepackage{color}
\usepackage{units}
\usepackage{slashed} 
\title{\boldmath \huge 750 GeV diphoton excess at CERN LHC from a dark sector assisted scalar decay}

\author[a]{Subhaditya Bhattacharya,}
\author[b]{Sudhanwa Patra,}
\author[c]{Nirakar Sahoo}
\author[c]{Narendra Sahu}
\affiliation[a]{Department of Physics, Indian Institute of Technology Guwahati, North Guwahati, Assam- 781039, India}
\affiliation[b]{Center of Excellence in Theoretical and Mathematical Sciences,
Siksha 'O' Anusandhan University, Bhubaneswar-751030, India}
\affiliation[c]{Department of Physics, Indian Institute of Technology Hyderabad, Kandi, Sangareddy, Medak 502 285, 
Telengana, India}

\emailAdd{subhab@iitg.ernet.in}
\emailAdd{sudhanwapatra@soauniversity.ac.in}
\emailAdd{nirakar.pintu.sahoo@gmail.com}
\emailAdd{nsahu@iith.ac.in}

\abstract{We present a simple extension of the Standard Model (SM) to explain the recent diphoton excess, reported by CMS and ATLAS at CERN LHC. 
The SM is extended by a dark sector including a vector-like lepton doublet and a singlet of zero electromagnetic charge, which are odd under a $Z_2$ symmetry. 
The charged particle of the vector-like lepton doublet assist the additional scalar, different from SM Higgs, to decay to di-photons of invariant mass around 750 
GeV and thus explaining the excess observed at LHC. The admixture of neutral component of the vector-like lepton doublet and singlet constitute the dark matter 
of the Universe. We show the relevant parameter space for correct relic density and direct detection of dark matter.}

\keywords{750~GeV Diphoton Excess, Dark Matter}
\begin{document}
\maketitle
\flushbottom
\section{Introduction}
\label{sec:intro}
Recently CMS and ATLAS detectors at the Large Hadron Collider (LHC) experiment\cite{lhcrun2a,atlasconf,CMS:2015dxe} reported an excess of $\gamma \gamma$ events 
in the proton-proton collision with centre-of-mass energy ($E_{cm}=\sqrt{s}$) 13 TeV. In fact, CMS reported the excess around 750 GeV with a local significance of 2.6 $\sigma$, 
while ATLAS reported the same excess around 750 GeV with a local significance of 3.6 $\sigma$ in the invariant mass distribution of $\gamma \gamma$. This excess could be simply 
due to the statistical fluctuations or due to the presence of a new Physics and needs future data for its verification. From ATLAS \cite{atlasconf} and CMS \cite{CMS:2015dxe} experiments, 
the production cross-section times the branching ratio of any resonance $X$ with a mass around $750$~GeV is given as:
\begin{align} 
&\sigma_{\rm ATLAS}\left( pp \to X \right) {\rm Br} \left( X\to \gamma \gamma \right) \simeq (10\pm 3) \mbox{fb}\, , \nonumber \\
&\sigma_{\rm CMS}\left( pp \to X \right) {\rm Br} \left( X\to \gamma \gamma \right) \simeq (6\pm 3) \mbox{fb}\,. \nonumber
\end{align}
The apparent conflict between these two experiments could be due to the different luminosity achieved even though the data are collected at the same centre-of-mass-energy  
$\sqrt{s}=13$ TeV. Amazingly the diphoton excesses observed by the two experiments are at the same energy bin. This gives enough indication for new physics 
beyond the standard model (SM) which can be confirmed or ruled out by future data. In the following we consider the diphoton excess observed at LHC to be a signature of 
new physics and provide a viable solution.   

If the diphoton events observed at LHC are due to a resonance, then the Landau-Yang's theorem~\cite{Landau:1948,Yang:1950rg} implies that
the spin of the resonance can not be 1. In other words the resonance could be a spin zero scalar or a spin two tensor similar to graviton. Another 
feature of the resonance is that the production cross-section times branching ratio is quite large ($\approx 10 {\rm fb}$), which indicates its production 
is due to strongly interacting particles. The most important feature of the resonance is that it's width is quite large ($\approx 45$ GeV). For large width 
of the resonance, the branching fraction to $\gamma \gamma$ events decreases significantly. Therefore, the main challenge for any theory beyond 
the SM is to find a large production cross-section: $\sigma \left( pp \to X \to \gamma \gamma \right)$ to fit the data. 

The diphoton signal around invariant mass of $750~$GeV can be explained via postulating a scalar resonance $S$ of $750~$GeV coupling 
to the vector like fermions arising in some new physics models. In such a situation, the diphoton signal can be reproduced by producing heavy 
scalar through gluon gluon fusion $gg\to S$ and calculating the branching fraction for the scalar decaying to two photons $\mbox{Br}(S\to \gamma \gamma)$. Although many 
attempts \cite{DiChiara:2015vdm,Pilaftsis:2015ycr,Knapen:2015dap,Backovic:2015fnp,Molinaro:2015cwg,Gupta:2015zzs,
Ellis:2015oso,Higaki:2015jag,Mambrini:2015wyu,Buttazzo:2015txu,Franceschini:2015kwy,Angelescu:2015uiz,
Bellazzini:2015nxw,McDermott:2015sck,Low:2015qep,Petersson:2015mkr,
Cao:2015pto,Kobakhidze:2015ldh,Agrawal:2015dbf,Chao:2015ttq,Fichet:2015vvy,Curtin:2015jcv,Csaki:2015vek,
Aloni:2015mxa,Demidov:2015zqn,No:2015bsn,Bai:2015nbs,Matsuzaki:2015che,Dutta:2015wqh,Becirevic:2015fmu,
Cox:2015ckc,Martinez:2015kmn,Bian:2015kjt,Chakrabortty:2015hff,Ahmed:2015uqt,
Deppisch:2016scs,Hernandez:2016rbi,Dutta:2016jqn,Modak:2016ung,
Danielsson:2016nyy,Chao:2016mtn,Csaki:2016raa,Karozas:2016hcp,
Ghorbani:2016jdq,Han:2016bus,Ko:2016lai,Nomura:2016fzs,
Ma:2015xmf,Palti:2016kew,Potter:2016psi,Jung:2015etr,
Marzola:2015xbh,Falkowski:2015swt,
Nakai:2015ptz,Harigaya:2015ezk,
Ellwanger:2015uaz,Karozas:00640,Csaki:00638,Chao:00633,Danielsson:00624,Ghorbani:00602,
Ko:00586,Han:00534,Nomura:00386,Palti:00285,Potter:00240,Palle:00618,Dasgupta:2015pbr,Bizot:2015qqo,
Goertz:2015nkp,Kim:2015xyn,Craig:2015lra,Cheung:2015cug,Allanach:2015ixl,Altmannshofer:2015xfo,Huang:2015rkj,
Belyaev:2015hgo,Liao:2015tow,Chang:2015sdy,Luo:2015yio,
Kaneta:2015qpf,Hernandez:2015hrt,Low:2015qho,Dong:2015dxw,Kanemura:2015vcb,Kanemura:2015bli,Kang:2015roj,Chiang:2015tqz,
Ibanez:2015uok,Huang:2015svl,Hamada:2015skp,Anchordoqui:2015jxc,Bi:2015lcf,Chao:2015nac,Cai:2015hzc,Cao:2015apa,
Tang:2015eko,Dev:2015vjd,Gao:2015igz,Cao:2015scs,Wang:2015omi,An:2015cgp,Son:2015vfl,Li:2015jwd,Salvio:2015jgu,
Park:2015ysf,Han:2015yjk,Hall:2015xds,Casas:2015blx,Zhang:2015uuo,Liu:2015yec,Das:2015enc,Davoudiasl:2015cuo,Cvetic:2015vit,
Chakraborty:2015gyj,Badziak:2015zez,Patel:2015ulo,Moretti:2015pbj,Gu:2015lxj,Cao:2015xjz,Pelaggi:2015knk,Dey:2015bur,
Hernandez:2015ywg,Murphy:2015kag,deBlas:2015hlv,Dev:2015isx,Boucenna:2015pav,Kulkarni:2015gzu,Chala:2015cev,
Bauer:2015boy,Cline:2015msi,Berthier:2015vbb,Kim:2015ksf,Bi:2015uqd,Heckman:2015kqk,Huang:2015evq,Cao:2015twy,Wang:2015kuj,
Antipin:2015kgh,Han:2015qqj,Ding:2015rxx,Chakraborty:2015jvs,Barducci:2015gtd,Cho:2015nxy,Feng:2015wil,Bardhan:2015hcr,
Han:2015dlp,Dhuria:2015ufo,Chang:2015bzc,Han:2015cty,Arun:2015ubr,Chao:2015nsm,Bernon:2015abk,Carpenter:2015ucu,Megias:2015ory,
Alves:2015jgx,Gabrielli:2015dhk,Kim:2015ron,Benbrik:2015fyz,Jiang:2015oms} have already been made in the context of a scalar resonance coupled 
to vector-like fermions, our prime goal in this paper is to show that the vector-like fermions assisting the production and decay of the scalar can be related 
to a possible dark sector. More precisely, we consider a dark sector including a vector-like lepton doublet $\psi$ and a singlet $\chi^0$, which are 
odd under a remnant $Z_2$ symmetry. The lightest $Z_2$ odd particle, which is an admixture of the neutral component of the lepton doublet $\psi$ and 
the singlet $\chi^0$, is postulated to constitute the dark matter (DM) component of the Universe. The charged particle in the lepton doublet $\psi$ on the other hand, 
assists the scalar resonance $S$ to decay to SM particles at one loop level. The decay of $S\to WW, ZZ,Z\gamma,\gamma\gamma,hh$ can easily enhance 
the width of the resonance, which is required to explain the width of the observed diphoton excess at LHC. Since the vector-like dark sector particles carry no color charges, they can not contribute to production of scalar particle via gluon fusion, {\it i.e.}, $gg \to S$, 
which is required to be large to fit the data. Hence, additional vector-like particles carrying color charges are introduced to aid the production (See for instance 
\cite{Chu:2012qy}). We then demonstrate the constraints on the model parameter space to explain $\gamma \gamma$ excess through  
$\sigma \left( pp \to S \to \gamma \gamma \right) \approx 10 ~{\rm fb}$ and dark sector phenomenology through relic density and direct search experiments. 

The paper is organized as follows: In Sec.\,\ref{model}, we present the model for $750~$GeV diphoton excess from
a dark sector assisted scalar decay. In Sec.\, \ref{diphoton_excess}, We discuss the diphoton signal from the decay 
of a scalar resonance which is primarily produced via gluon-gluon fusion process at LHC. We then present the relic density 
and direct detection constraints on DM parameter space in Sec.\,\ref{relic_density}, which is consistent with
$750~$GeV diphoton excess and finally conclude in Sec.\,\ref{conclusions}.

\section{Model for dark sector assisted diphoton excess} \label{model}

We extend the SM with a scalar singlet $S(1,1,0)$ and a dark sector, comprising of a vector like lepton doublet 
$\psi^T=(\psi^0, \psi^-)$ (1,2,-1) and a leptonic singlet $\chi^0$ (1,1,0), where the quantum numbers in the parentheses are 
under the gauge group $SU(3)_c\times SU(2)_L\times U(1)_Y$. In addition to the SM gauge symmetry, we impose a discrete symmetry 
$Z_2$ under which the dark sector fermions: $\psi$ and $\chi^0$ are odd, while all other fields are even. The motivation for 
introducing such a dark sector is two fold: i) firstly, the linear combination of the neutral component of the lepton doublet ($\psi^0$)
and singlet ($\chi^0$) becomes a viable candidate of DM, ii) secondly, the charged component of the vector like lepton doublet 
assists the scalar resonance $S$ to give rise the diphoton excess of invariant mass $750~$ GeV.

The relevant Lagrangian can be given as:
\begin{eqnarray}
\label{lagrangian}
-\mathcal{L} &&\supset M_\psi \overline{\psi} \psi + f_\psi S \overline{\psi} \psi + M_\chi \overline{\chi^0} \chi^0 
+ f_\chi S \overline{\chi^0}\chi^0 \nonumber \\
&&+ \left[ Y\overline{\psi}\widetilde{H}\chi^0 + {\rm h.c.}\right] + V(S,H)\,,
\end{eqnarray}
where $H$ is the SM Higgs isodoublet and $\widetilde{H}=i\tau_2 H^*$. The scalar potential in Eq. (\ref{lagrangian}) is given by 
\begin{eqnarray}\label{scalarpotential}
V(S,H) &&= \mu_H^2 H^\dagger H + \lambda_H (H^\dagger H)^2 + \frac{1}{2} \mu_S^2 S^2 + \frac{\lambda_S}{4} S^4 \nonumber \\
&&+ \frac{\lambda_{SH}}{2}(H^\dagger H) S^2 + \mu_{SH} S H^\dagger H\,,
\end{eqnarray}
where $\lambda_H, \lambda_S > 0$ and $\lambda_{SH} > -2\sqrt{\lambda_S \lambda_H}$ is required for vacuum stability. 
We assume that $\mu_S^2 > 0$ and $\mu_H^2 < 0$, so that $S$ does not acquire a vacuum expectation value (vev) before 
electroweak phase transition. After $H$ acquires a vev: $\langle H \rangle = v=\sqrt{-\mu_H^2/2\lambda_H}$, $S$ gets an induced vev which we neglect in the following calculation. 

\begin{figure}[!h]
	\centering
	\includegraphics[width=0.52\textwidth]{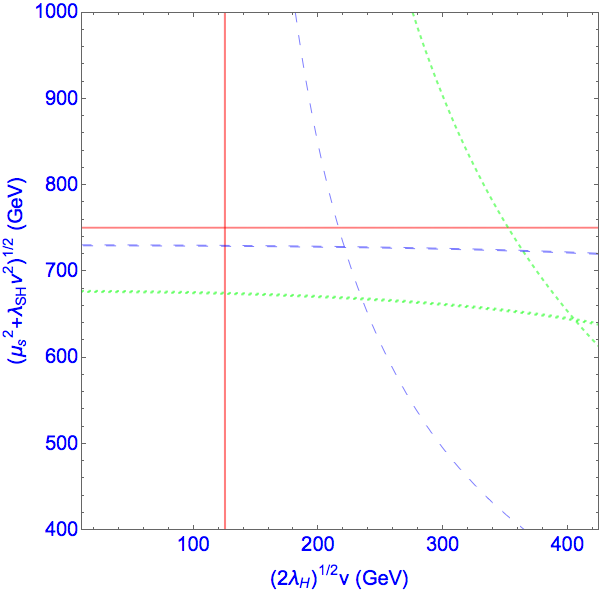}
	\caption{Contours of $M_h=125~ {\rm GeV}$ and $M_S=750 ~{\rm GeV}$ in the plane of $\sqrt{ 2 \lambda_H} v$ and $\sqrt{ \mu _S^2 + \lambda_{SH}v^2}$ for   
	 $\mu_{SH}= 10$~GeV (Solid red), $\mu_{SH}=750$~GeV (Dashed blue), and $\mu_{SH}=1400$~GeV (Dotted green).}
	\label{hs_plot}
\end{figure}

After electroweak phase transition, $S$ mixes with the $H$ through the tri-linear term $S H^\dagger H$. Due to the mixing 
we get the mass matrix for the scalar fields as: 
\begin{equation}\label{mass_matrix}
\mathcal{M}^2=\begin{pmatrix} 2\lambda_H v^2 & \mu_{SH}  v \cr\\
\mu_{SH}v &  \mu_S^2 + \lambda_{SH} v^2 \end{pmatrix}\,,
\end{equation}
where the trilinear parameter $\mu_{SH}$ (with mass dimension one) decides the mixing between the two scalar fields, which can be parameterized by 
a mixing angle $\theta_{hS}$ as
\begin{equation}
\tan \theta_{hS} = \frac{\mu_{SH} v}{\mu_S^2 + \lambda_{SH}v^2 - 2\lambda_H v^2}\,.
\end{equation} 
The above equation shows that the mixing angle $\theta_{hS}$ between the two scalar fields vanishes if $\mu_{SH} \to 0$. 
For finite mixing, the masses of the physical Higgses can be obtained by Diagonalizing the mass matrix (\ref{mass_matrix}) 
and is given by:
\begin{eqnarray}\label{mass}
M_h^2 &=& \left( \lambda_H v^2 + \frac{1}{2}\mu_S^2+ \frac{1}{2}\lambda_{SH} v^2 \right) + \frac{1}{2} D \nonumber\\ 
M_S^2 &=& \left( \lambda_H v^2 + \frac{1}{2}\mu_S^2 + \frac{1}{2} \lambda_{SH} v^2 \right) - \frac{1}{2} D \,,
\end{eqnarray}
where $D=\sqrt{\left(2\lambda_H v^2 -\mu_S^2-\lambda_{SH}v^2  \right)^2 + 4 \left(\mu_{SH} v \right)^2} $, corresponding to the mass 
eigenstates $h$ and $S$, where we identify $h$ as the SM Higgs with $M_h=125$ GeV and S is the new scalar with $M_S= 750$ GeV. Using Eq.\,(\ref{mass}) 
we have plotted contours for $M_h=125 ~{\rm GeV}$ and $M_S=750 ~{\rm GeV}$ in the plane of $\sqrt{ 2\lambda_H} v$ and $\sqrt{ \mu _S^2 + \lambda_{SH}v^2}$ 
for different choices of $\mu_{SH}=\{10, 750, 1400\}~$GeV (Red thick, Blue dashed and Green dotted lines respectively), as shown in Fig.\,\ref{hs_plot}. 
We observe that for small mixing ($\mu_{SH}=10$ GeV, represented by red solid line) contours of $M_S=750~$GeV and $M_h=125~$GeV intersect vertically 
as expected while for larger mixing, $\mu_{SH}>1400~$ GeV, we can not get simultaneous solution for $M_S=750~$GeV and $M_h=125~$GeV. This implies 
that the largest allowed mixing for which we get the simultaneous solution is $\sin \theta_{hS} \approx 0.467$. However, such large values of the mixing angles 
are strongly constrained from other observations (See for 
instance \cite{Falkowski:2015swt}). We will get back to this issue of small/large mixing in Sec.\,\ref{diphoton_excess}.  

The electroweak phase transition also gives rise a mixing between $\psi^0$ and $\chi^0$. In the basis $(\chi^0, \psi^0)$, the mass matrix 
is given by:
\begin{equation}
\mathcal{M} = \begin{pmatrix}  M_\chi & m_D\cr \\
m_D & M_\psi \,,
\end{pmatrix}
\end{equation} 
where $m_D=Yv$. Diagonalizing the above mass matrix we get the mass eigen values as
\begin{eqnarray}
M_1 \approx M_\chi + \frac{m_D^2}{M_\psi - M_\chi}\nonumber\\
M_2 \approx M_\psi - \frac{m_D^2}{M_\psi - M_\chi}
\end{eqnarray}
where we have assumed $m_D << M_\psi, M_\chi$. The corresponding mass eigenstates are given by 
\begin{eqnarray}
\psi_1= \cos \theta \chi^0 + \sin \theta \psi^0\nonumber\\
\psi_2=\cos \theta \psi^0 - \sin \theta \chi^0\,,
\end{eqnarray}
where the mixing angle is given by:
\begin{equation}
\tan 2\theta = \frac{2 m_D}{M_\psi - M_\chi}\,. 
\end{equation}
in small mixing limit. We assume that $\psi_1$ is the lightest odd particle and hence constitute the DM of the Universe. Note that 
$\psi_1$ is dominated by the singlet component with a small admixture of doublet, while $\psi_2$ is dominantly a doublet with a small admixture 
of singlet component. This implies that $\psi_2$ mass is required to be larger than 45 GeV in order not 
to conflict with the invisible Z-boson decay width. In the physical spectrum we also have a charged fermion $\psi^\pm$ whose mass in terms 
of the masses of $\psi_{1,2}$ ($M_{1,2}$) and the mixing angle $\theta$ is given by
\begin{equation}
M^\pm = M_1 \sin^2 \theta + M_2 \cos^2 \theta 
\end{equation}
In the limit of vanishing mixing in the dark sector, $\sin \theta \to 0$, $M^\pm = M_\psi$. Therefore, a non-zero mixing also gives rise to a mass splitting between 
$\psi^\pm$ and $\psi_2$ is given by $\Delta M  = \frac{m_D^2}{M_\psi - M_\chi}$.

The fermions interact to the SM gauge bosons through the following interaction terms:

\begin{eqnarray}
\label{gauge-int}
-\mathcal{L}_{W} \supset &&  \frac{g\sin \theta}{\sqrt{2}} 
\overline{\psi_1}\gamma^\mu W_\mu^+ \psi^- + \frac{g\cos \theta}{\sqrt{2}} \overline{\psi_2}\gamma^\mu W_\mu^+ \psi^-\,, \nonumber \\
-\mathcal{L}_{Z} \supset && \frac{g}{2\cos \theta_w} \left( \sin^2 \theta \overline{\psi_1} \gamma^\mu Z_\mu \psi_1 
+ \sin \theta \cos \theta ( \overline{\psi_1} \gamma^\mu Z_\mu \psi_2 + \overline{\psi_2} \gamma^\mu Z_\mu \psi_1) + \cos^2 \theta \overline{\psi_2} \gamma^\mu Z_\mu \psi_2 \right) \, \nonumber \\ 
&& + ~\frac{g}{2}\overline{\psi^-} \gamma^\mu Z_\mu \psi^-\,, \nonumber \\
-\mathcal{L}_{\gamma} \supset && e\gamma^\mu \overline{\psi^-} A_\mu \psi^- \,.
\end{eqnarray}

\section{Explanation for diphoton Excess}\label{diphoton_excess}
\subsection{$S\to\gamma \gamma$ and production of $S$ through mixing with the SM Higgs}\label{s-h-mixing}
The LHC search strategy for diphoton events, if possible via a scalar resonance $S$ with mass around 
$750$~GeV, is mostly decided by the production and subsequent decay of the resonant particle to $\gamma \gamma$, 
which can be parametrized as:
\begin{equation}
\sigma_{\rm ATLAS/CMS} \left(pp \to S \to \gamma \gamma \right) \simeq \sigma_{\rm prod}\left(pp \to S \right) 
\cdot \mbox{Br.}\left(S \to \gamma \gamma \right)\,. 
\end{equation}
The above cross-section has to be compared with the experimental data
\begin{align}\label{expt_value}
&\sigma_{\rm ATLAS}\left( pp \to X \to \gamma \gamma \right) \simeq (10\pm 3) \mbox{fb}\, ,\\
&\sigma_{\rm CMS}\left( pp \to X \to \gamma \gamma \right) \simeq (6\pm 3) \mbox{fb}\,.
\end{align}
In absence of the additional vector-like fermions, the production of $S$ and its subsequent decay to $\gamma$ $\gamma$ 
can occur through the mixing with the SM Higgs, which can be given as:   
\begin{equation}
\sigma (pp \to S \to \gamma \gamma) \simeq \sigma_{\rm prod}\left(pp \to h \right) \cdot \sin^4\theta_{hS} \cdot 
\frac{\Gamma\left(h \to \gamma \gamma \right)}{\Gamma(S \to \mbox{All})}\,,
\end{equation}
where $\Gamma(S \to \mbox{All}) \approx 45$ GeV as indicated by ATLAS data \cite{atlasconf}. Within the SM, the decay width: 
$h \to \gamma \gamma $ can be estimated to be $\approx 4 \times 10^{-6} {\rm GeV} $ for $M_h=125$ GeV and 
$\Gamma( h \to \mbox{All})= 4~ {\rm MeV}$. The total production cross-section of Higgs at centre of mass energy of 13 TeV is given 
by $\approx 50 ~{\rm pb}$ ~\cite{ATLAS_CONFERENCE}. Thus with a maximal mixing between the SM Higgs and S, {\it i.e.} $(\sin \theta_{hS}\approx 0.4$, we see that $\sigma \left(pp \to S \to \gamma \gamma \right) \approx 10^{-4} {\rm fb}$, 
which is much smaller than the required value given in Eq. (\ref{expt_value}). Therefore, we conclude that the production of the scalar resonance $S$ giving diphoton excess at LHC can not be possible through its mixing with the SM Higgs. 

In the following sub-section\,\ref{ggtoS} we set the $S-h$ mixing to be zero and adopt an alternative scenario for $\sigma \left(pp \to S \to \gamma \gamma \right)$ using vector-like quarks.

\subsection{Dark sector assisted $S$ decays}\label{Stogamma}
Since $S$ is a singlet scalar, it can not directly couple to the gauge bosons. On the other hand, $S$ can couple to vector-like dark sector fermions 
which can couple to SM gauge bosons as discussed in the previous section. As the charged component of the $Z_2$ -odd fermion doublet assist the decay of $S$, 
we term it as dark sector assisted decay. Defining $B_{\mu \nu}$ and $W^i_{\mu \nu}$ as the respective field strength tensors for 
the gauge group $U(1)_{\rm Y}$ and $SU(2)_{L}$, one can write down the effective operators for coupling between the scalar $S$ and the 
vector bosons by integrating out the vector-like fermions in the loop as:
\begin{align}
\mathcal{L}_{\rm EFT}& \supset
  \kappa_2 S W^i_{\mu \nu} W^{i,\mu \nu} +\kappa_1 S B_{\mu \nu} B^{\mu \nu} 
\end{align}
where the effective couplings $\kappa_1$ and $\kappa_2$ can be expressed in terms of Yukawa coupling $f_\psi$ connecting 
scalar with vector-like fermion $\psi$ as ~\cite{Dutta:2015wqh}: 
\begin{equation}
k_1=\frac{f_\psi g_Y^2}{32 \pi^2 M_\psi} \,\,\, {\rm and} \,\,\, 
k_2=\frac{3f_\psi g^2}{64 \pi^2 M_\psi}
\end{equation}
Since the vector-like dark sector particles carry no color charge and hence, can not contribute to the decay of $S\to gg$ and $gg \to S$ for production of scalar particle. However, one can produce large cross-section for scalar $S$ via gluon fusion process by introducing additional vector-like particle carrying color charge, for example, see ref.~\cite{Ellis:2015oso,McDermott:2015sck}. We will also adopt a similar strategy that will be discussed in the next sub-section. After rotation to the physical gauge boson 
states the decay rates can be given as:
\begin{eqnarray}
\Gamma \left(S \rightarrow WW\right) &=& \frac{1}{16 \pi} \left[2+\left(1-\frac{M_S^2}{2M_W^2}\right)^2\right]\nonumber\\ 
&& \left(1-\frac{4 M_W^2}{M_S^2}\right)^{1/2}k_{WW}^2 M_S^3 \nonumber \\
\Gamma \left(S \rightarrow ZZ\right) &=& \frac{1}{32 \pi} \left[2+\left(1-\frac{M_S^2}{2M_Z^2}\right)^2\right]\nonumber\\ 
&& \left(1-\frac{4 M_Z^2}{M_S^2}\right)^{1/2} k_{ZZ}^2 M_S^3 \nonumber\\
\Gamma \left(S \rightarrow Z\gamma \right) &=& \frac{3}{16 \pi}  \left(1-\frac{M_Z^2}{M_S^2}\right) k_{Z\gamma}^2 M_S^3 \nonumber\\
\Gamma \left(S \rightarrow \gamma \gamma \right) &=& \frac{1}{8 \pi} k_{\gamma \gamma}^2 M_S^3 \nonumber \\
\end{eqnarray}
where the effective couplings are given by \cite{Dutta:2015wqh}: 
\begin{eqnarray}\label{effective_couplings}
k_{WW} &=& \frac{g^2}{32\pi^2} \frac{f_\psi}{M_\psi} A_{1/2}(x_\psi)\nonumber\\
k_{\gamma \gamma} &=& \frac{e^2}{16\pi^2} Q_\psi^2 \frac{f_\psi}{M_\psi} A_{1/2}(x_\psi)\nonumber\\ 
k_{ZZ} &=& k_{WW} (1- \tan^2 \theta_W) + k_{\gamma \gamma} \tan^2 \theta_W\nonumber\\
k_{Z\gamma} &=& k_{WW} \cos 2\theta_W \tan \theta_W - k_{\gamma \gamma} 2 \tan \theta_W \nonumber \\
\end{eqnarray} 
The factors involved in Eq.\,(\ref{effective_couplings}) are given by
\begin{align}\label{loop_function}
	&A_{1/2}(x_\psi) = 2x_\psi \left[ 1 + (1- x_\psi ) f(x_\psi) \right]\, , \nonumber \\
	&x_\psi = \frac{ 4 M_\psi^2}{M_S^2} \, \, , \nonumber \\
	&f(x) = \left\{ \begin{matrix}\arcsin^{2} \sqrt{x} & & x \leq 1 \\ 
	    -\frac{1}{4}\left[ {\rm ln} \left( \frac{1+\sqrt{1-x}}{1-\sqrt{1-x}} \right)-i\pi\right]^2 
			&& x\geq 1.\end{matrix} \right.
\end{align}

\subsection{Dark sector assisted $S\to \gamma \gamma$ and quark-like vector particles for $gg\to S$}\label{ggtoS}
\begin{figure}[!h]
	\centering
	\includegraphics[width=0.52\textwidth]{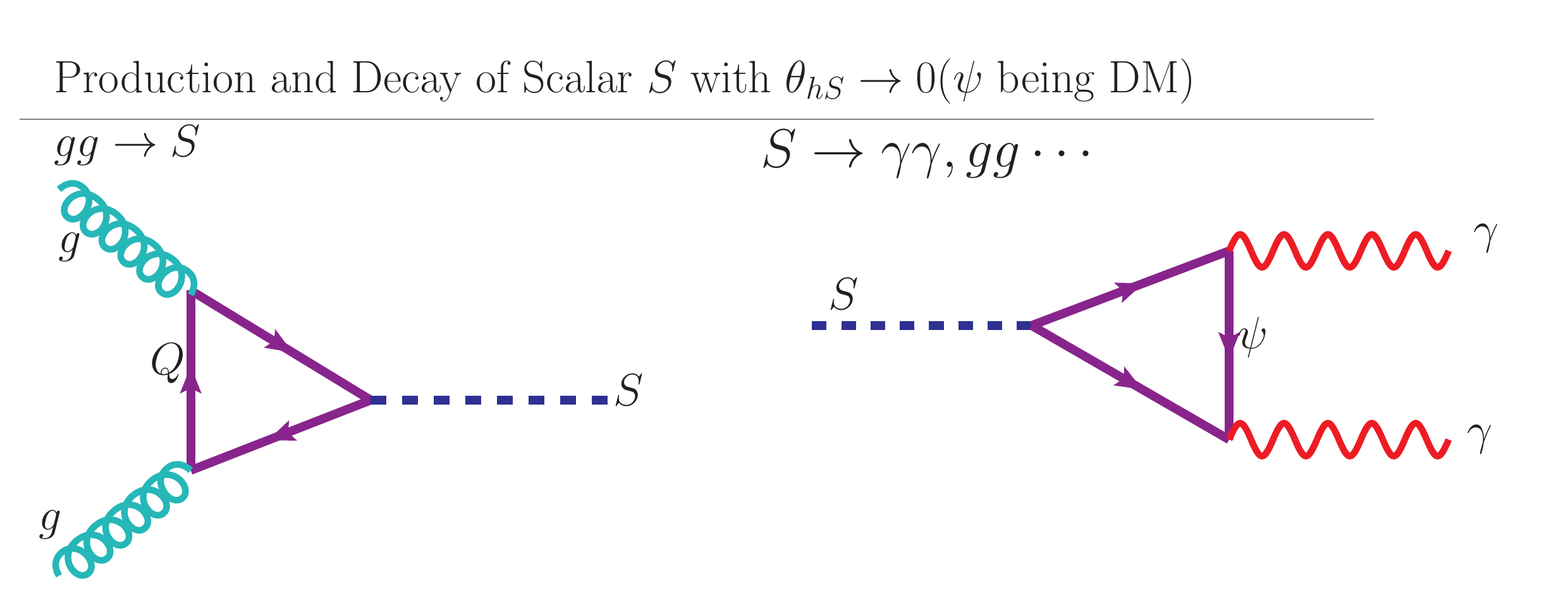}
	\caption{Feynman diagrams for production of scalar $S$ through gluon gluon fusion mediated by quark-like vector particle $Q$ and its subsequent decay to SM particles mediated 
	by the dark sector particle $\psi^\pm$. The other decay modes of $S$ via its mixing with the SM Higgs are suppressed in the limit $\theta_{hS} \to 0$.}
\label{feyn:dmassist2}
\end{figure}

As discussed in section\,\ref{s-h-mixing}, we see that the required cross-section for the scalar resonance $S$ production can not be achieved through S-h mixing. 
As an alternative, we introduce an iso-singlet quark-like vector fermion $Q$ of mass $M_Q$ to the framework discussed in the above section. The main reason for introducing 
additional quark-like vector particle is to provide the large production cross-section for scalar $S$ via gluon gluon fusion process as shown in the left-panel of Fig.\,\ref{feyn:dmassist2} 
even with $\theta_{hS} \to 0$. The subsequent decay $S\to \gamma \gamma$ mediated by $\psi^\pm$ is shown in right-panel of Fig.\,\ref{feyn:dmassist2}. We note that the additional 
vector-like quark also plays an important role in the relic density of DM as we shall discuss in section\,\ref{relic_density}. 

The Yukawa coupling of the scalar $S$ to $Q$ can be given as $f_Q S \bar{Q} Q$. This coupling helps in 
producing $S$ via gluon gluon fusion process. The production of scalar $S$, arising from gluon gluon fusion process, 
and its subsequent decay to $\gamma \gamma$ can be expressed in terms of the decay rate $\Gamma(S\to gg)$ 
as \cite{Djouadi:2005gi,Franceschini:2015kwy}:
\begin{equation}\label{stogg}
\sigma (pp \to S \to \gamma \gamma) = \frac{1}{M_S \hat{s} } C_{gg} \Gamma(S\to gg) {\rm Br} (S\to \gamma \gamma)
\end{equation}
where $\sqrt{\hat{s}}= 13$ TeV is the centre of mass energy at which LHC is collecting data. The dimensionless coupling 
\begin{equation}
C_{gg} = \frac{\pi^2}{8} \int_{M_S^2/\hat{s}}^{1} \frac{dx}{x} g(x) g(M_S^2/\hat{s} x)\,.
\end{equation}
At $\sqrt{\hat{s}}= 13$ TeV, $C_{gg}=2137$~\cite{Franceschini:2015kwy}. In Eq. (\ref{stogg}), the decay rate 
$\Gamma(S\to gg)$ is given by:
\begin{equation}
\Gamma \left(S \rightarrow  gg \right) = \frac{1}{8 \pi} k_{gg}^2 M_S^3
\end{equation}
where the effective coupling of $S$ to $gg$ through the exchange of Q in the loop is given by 
\begin{equation}
k_{gg} = \frac{g_S^2}{16 \pi^2} \frac{f_Q}{M_Q}N_c A_{1/2}(x_Q)
\end{equation}
where $A_{1/2}(x)$ is given by Eq. (\ref{loop_function}). 

\begin{figure}[!h]
	\centering
	\includegraphics[width=0.5\textwidth]{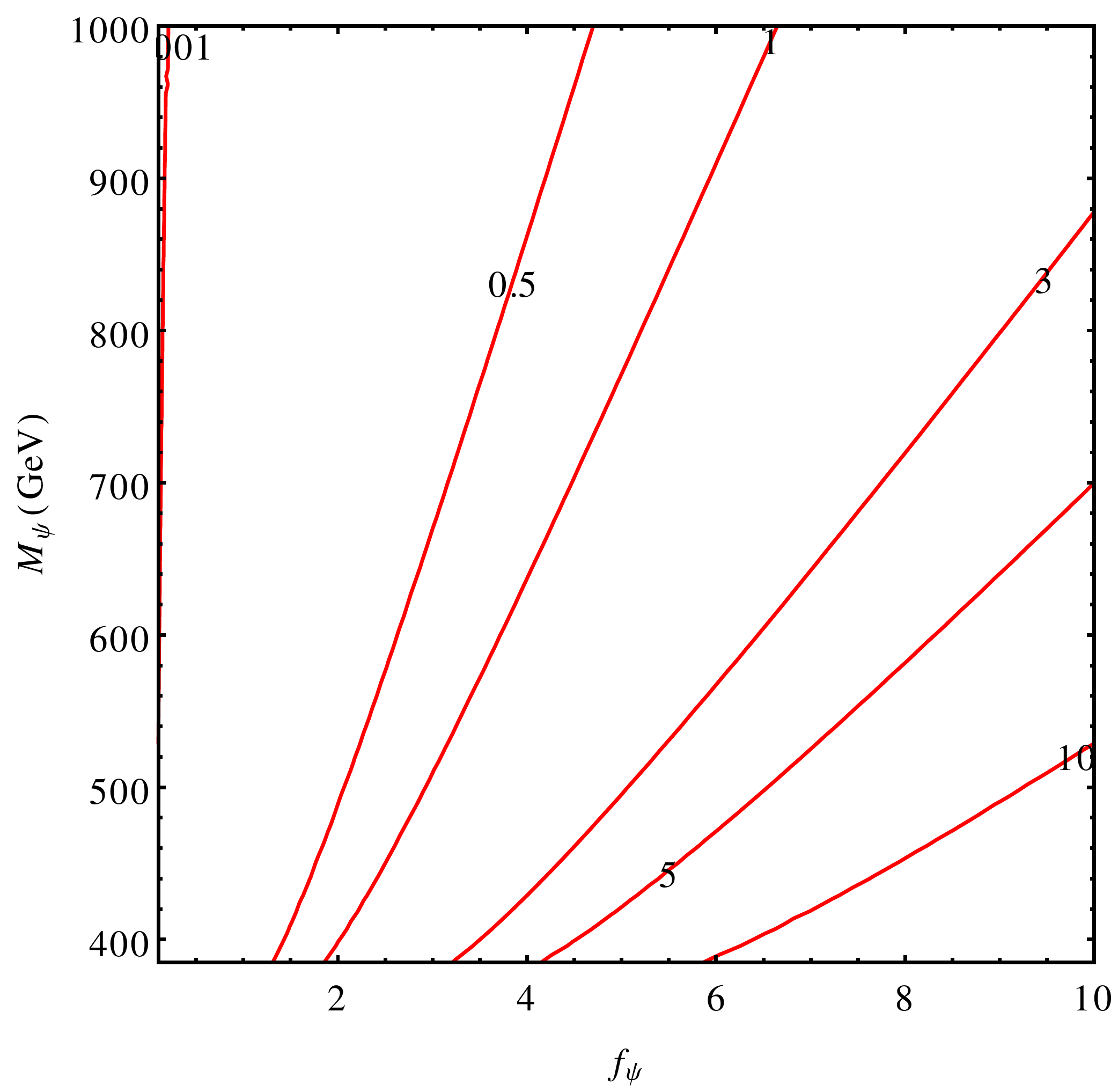}
	\caption{Contours of $\sigma (pp \to S \to \gamma \gamma)$ in the plane of $f_\psi$ versus $M_\psi$ for 
                 $f_\psi=f_Q$, $M_\psi=M_Q$ and  $\sin \theta_{hS}=0$.}
	\label{production_contour}
\end{figure}
\begin{figure}[!h]
	\centering
	\includegraphics[width=0.5\textwidth]{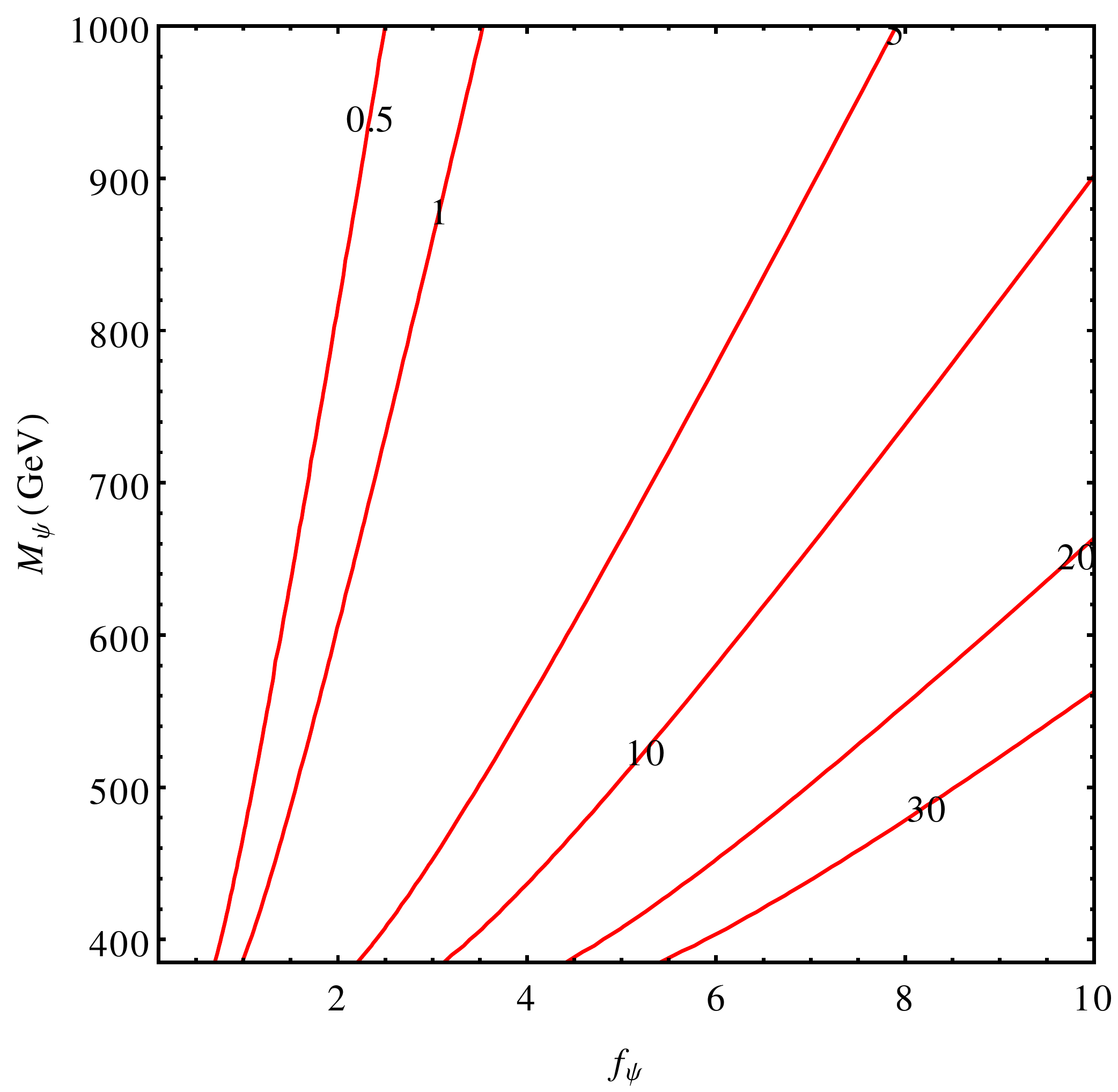}
	\caption{Contours of $\Gamma(S \rightarrow {\rm All})$ (in GeV)in the plane of $f_\psi$ versus $M_\psi$ for 
                 $f_\psi=f_Q$, $M_\psi=M_Q$ and $\sin \theta_{hS}=0$.}
	\label{totdecaywidth}
\end{figure}  
\begin{figure}[!h]
	\centering
	\includegraphics[width=0.5\textwidth]{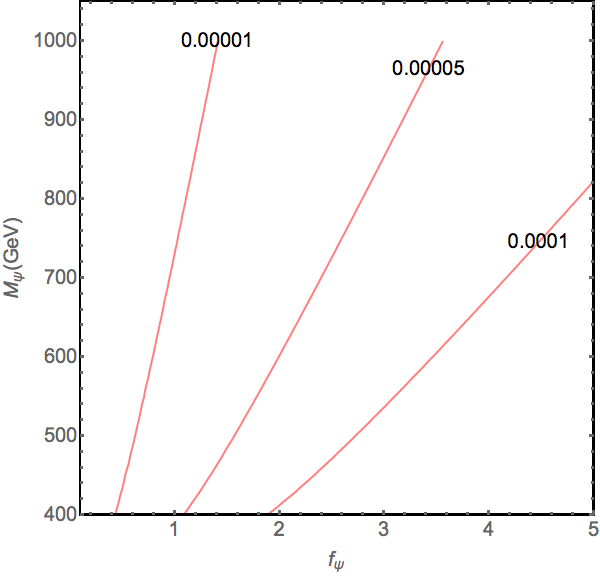}
	\caption{Contours of {\rm Br} ($S \rightarrow \gamma \gamma$) in the plane of $f_\psi$ versus $M_\psi$ for 
                 $f_Q=5$ and $M_Q=600$ GeV. We set $\sin \theta_{hS}=0$.}
	\label{branchingratio}
\end{figure}
In Fig. (\ref{production_contour}), we have shown the contours of $\sigma (pp \to S \to \gamma \gamma)$ in the plane of 
$f_\psi$ versus $M_\psi$ by assuming that $f_\psi = f_Q$ and $M_\psi= M_Q$. From Fig. (\ref{production_contour}), we see that to get a production cross-section 
of 10 fb, we need the $S$ coupling to $f_\psi = f_Q > 5$ . The mass of these vector-like fermions are chosen to be larger than 375 GeV in order to avoid the tree level 
decay of $S \to \bar{Q}Q$. The corresponding total decay width and branching fraction are shown in Fig. (\ref{totdecaywidth}) and (\ref{branchingratio}). We see that 
the total decay width can be as large as 30 GeV, while the branching fraction is order of $10^{-4}$. Since the mass of the vector-like 
fermions are heavier than 375 GeV, the decay of $S$ to SM particles occurs via the triangle loop constituting $\psi^\pm$. However, 
the tree level decay of $S$ to $hh$ is allowed. It may increase the total width depending on the mixing between SM Higgs and S. However, 
we have checked that for $\sin \theta_{hS} < 0.1$, the tree level decay of $S$ to $hh$ does not affect the above result.

\section{Relic Density and Direct Search constraints on Dark Matter} \label{relic_density}
In the previous sections we discussed the role of charged component $(\psi^\pm)$ of the leptonic doublet $\psi$ in the diphoton excess. Now 
we will show that the neutral component of $\psi$, i.e. $\psi_0$, and $\chi^0$ combine to explain the relic abundance of DM. As discussed in Section \ref{model}, 
we use $\psi_1$ of mass $M_1$ and $\psi_2$ of mass $M_2$, which are linear combination of the states $\psi_0$, and $\chi^0$. We assume that $\psi_1$ which 
is the lightest $Z_2$ odd particle, which is dominated by a singlet component $\chi^0$, constitutes the DM of the Universe. The relic density of the DM is 
mainly dictated by annihilations $\overline{\psi_1}\psi_1 \to W^+W^-$ and $\overline{\psi_1} \psi_1 \to h h$ through $Z$ and Higgs mediation. The other relevant channels are mainly co-annihilation of $\psi_1$ with $\psi_2$ and $\psi^\pm$. For details see ref. \cite{Bhattacharya:2015qpa,Cynolter:2008ea,Cohen:2011ec,Cheung:2013dua}. However, this particular model with additional vector like quarks ($Q$) marks a significant departure through annihilations of the DMs to vector-like quarks, 
$\overline{\psi_1} \psi_1 \to \bar{Q}Q$ through $S$ mediation. Due to the large couplings required to explain the observed diphoton excess, the annihilations to vector like quarks dominate over the others whenever $M_{DM} \ge M_Q$ and the relic density diminishes significantly in those regions irrespective of the other parameters.

\begin{figure}[t!]
$$
\includegraphics[scale=0.35]{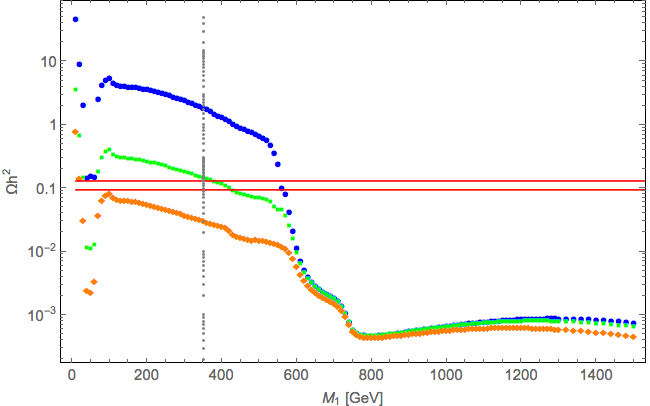}
\includegraphics[scale=0.35]{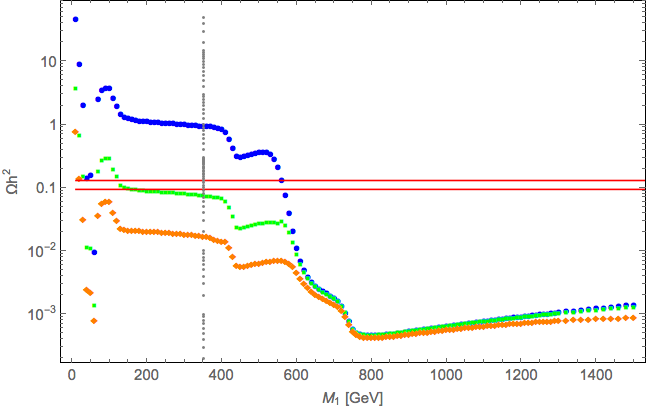}
$$
\caption{Variation of relic density ($\Omega h^2$) with  DM mass ( $M_1$ in GeV) for $M_Q=600$ GeV. $\sin \theta=\{0.1,0.2,0.3\}$ cases (from top to bottom) 
are depicted simultaneously in blue green and orange. On the left panel we have taken $\Delta M\equiv M_{2}-M_{1}=$ 100 GeV while on the 
right panel we have set $\Delta M\equiv M_{2}-M_{1}=$ 500 GeV. Red band indicates relic density within WMAP range. The region to the right side of 
vertical dotted line denotes compatibility with 750 GeV diphoton excess.}
\label{fig:Omegam-1}
\end{figure}
\begin{figure}[t!]
$$
\includegraphics[scale=0.35]{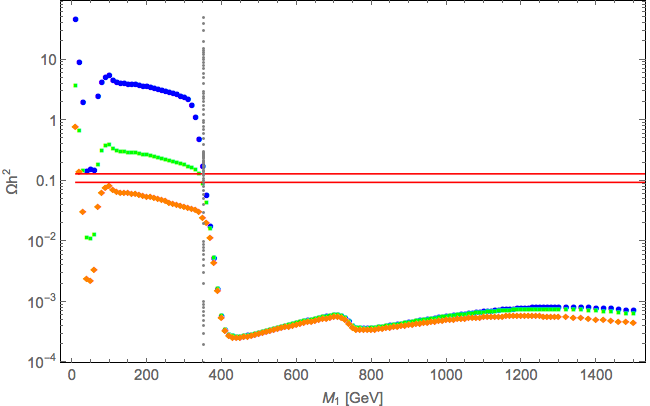}
\includegraphics[scale=0.35]{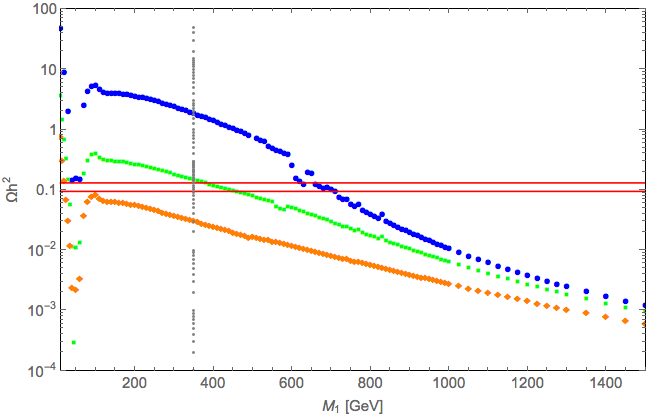}
$$
\caption{Variation of relic density ($\Omega h^2$) with  DM mass ( $M_1$ in GeV) for $M_Q= 400$ GeV (left) and $M_Q> 2000$ GeV (right). $\sin \theta=\{0.1,0.2,0.3\}$ 
cases (from top to bottom) are depicted simultaneously in blue green and orange. On both panels we have taken $\Delta M\equiv M_{2}-M_{1}=$ 100 GeV. Red band indicates 
relic density within the WMAP range. The region to the right side of vertical dotted line denotes compatibility with 750 GeV excess.}
\label{fig:Omegam-2}
\end{figure}

\begin{figure}[h!]
$$
\includegraphics[scale=0.40]{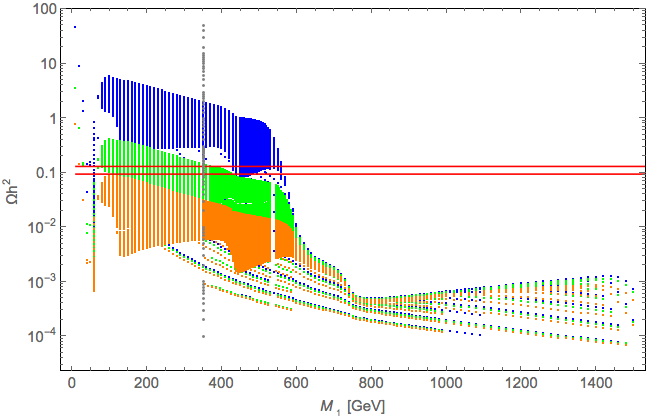}
$$
\caption{Same as Fig. \ref{fig:Omegam-1}, but with all possible $\Delta M$. $M_Q=~600$ GeV.}
\label{fig:Omegam-3}
\end{figure}

The relic density of the $\psi_1$ DM can be given by~\cite{griest}. 
\begin{equation}
\Omega  h^2 = \frac{1.09 \times 10^9 \rm GeV^{-1}}{g_\star ^{1/2} M_{PL}} \frac{1}{J(x_f)}\,,
\end{equation}
where $J(x_f)$ is given by
\begin{equation}
J(x_f)= \int_{x_f}^ \infty \frac{\langle \sigma |v| \rangle _{eff}}{x^2} \hspace{.2cm} dx \,,
\end{equation}
where $\langle \sigma |v| \rangle _{eff}$ is the thermal average of DM annihilation 
cross sections including contributions from co annihilations as follows:
\begin{equation}\label{sigma_eff}
\begin{split}
\langle \sigma |v| \rangle _{eff} = & \frac{g_1^2}{g_{eff}^2} \sigma (\psi_1 \psi_1)
 +2 \frac{g_1 g_2}{g_{eff}^2} \sigma (\psi_1 \psi_2) (1+\Delta)^{3/2} exp(-x\Delta) \\
 & +2 \frac{g_1 g_3}{g_{eff}^2} \sigma (\psi_1 \psi^-) (1+\Delta)^{3/2} exp(-x\Delta)\\
 &+2 \frac{g_2 g_3}{g_{eff}^2} \sigma (\psi_2 \psi^-) (1+\Delta)^{3} exp(-2x\Delta)\\
& + \frac{g_2 g_2}{g_{eff}^2} \sigma (\psi_2 \psi_2) (1+\Delta)^{3} exp(-2x\Delta) \\
& + \frac{g_3 g_3}{g_{eff}^2} \sigma (\psi^- \psi^-) (1+\Delta)^{3} exp(-2x\Delta)\,.
\end{split}
\end{equation}
In the above equation $g_1$,$g_2$ and $g_3$ are the spin degrees of freedom for $\psi_1$, $\psi_2$ and $\psi^-$ respectively. Since these are spin half 
particles, all g's are 2. The freeze-out epoch of $\psi_1$ is parameterized by $x_f= \frac{M_{1}}{T_f}$, where $T_f$ is the freeze out temperature. 
$\Delta$ depicts the mass splitting ratio as $\Delta = \frac{M_{i}- M_{1}}{ M_{1}} $, where $M_i$ stands for the mass of 
$\psi_2$ and $\psi^{\pm}$. The effective degrees of freedom $g_{eff}$ in Eq. (\ref{sigma_eff}) is given by
 \begin{equation}
 g_{eff} = g1+ g_2 (1+\Delta)^{3/2} exp(-x\Delta) + g_3 (1+\Delta)^{3/2} exp(-x\Delta)\,.
\end{equation}  

The dark-sector, spanned by the $Z_2$ odd vector-like fermions, is mainly dictated by the following three parameters : 
\begin{equation}\label{parameter-space}
\sin\theta, M_{1}, M_{2}\,.
\end{equation}
In the following we shall vary the parameters in Eq. (\ref{parameter-space}) and find the allowed region of correct relic abundance 
for $\psi_1$ DM satisfying WMAP~\cite{Hinshaw:2012aka} constraint ~\footnote{The range we use corresponds to the WMAP results; the PLANCK 
constraints $0.112 \leq \Omega_{\rm DM} h^2 \leq 0.128$~\cite{Ade:2013zuv}, though more stringent, do not lead to significant changes in 
the allowed regions of parameter space.}
\begin{equation}
0.094 \leq \Omega_{\rm DM} h^2 \leq 0.130 \,.
\label{eq:wmap.region} 
\end{equation}

A notable set of parameters that also crucially controls the allowed DM parameter space are the vector like quark masses ($M_Q$), the scalar mass ($M_S$), 
the couplings of the vector-like quarks to $S$ ($f_Q$) and coupling of $S$ to the dark sector ($f_\psi$) due to the annihilation of DMs to vector like quarks 
($\overline{\psi_1} \psi_1 \to \bar{Q}Q$) through $S$ mediation: 

\begin{equation}\label{parameter-space2}
 M_S, M_{Q}, f_\psi, f_Q\,.
\end{equation}

An interesting DM phenomenology is likely to evolve with an arbitrary variation of these parameters depending on which this particular annihilation channel compete 
with others. However, given that one of the primary goals of this model is to explain the diphoton excess, in the following scans we choose a set of specific values 
for these parameters, that are required to explain the collider signature, as:

\begin{equation}\label{parameter-space2}
 M_S= 750 ~{\rm GeV}, M_{Q}= 600 ~{\rm GeV}, f_\psi =f_Q=5\,.
\end{equation}
Note that vector-like quark masses lighter than 600 GeV are strongly constrained by the direct searches at collider~\cite{Agashe:2014kda}. For $M_Q > 600$ GeV, we need
even larger couplings to explain the observed diphoton excess. We note here although the large couplings required are within the perturbative limit ($f_\psi, f_Q < 4 \pi$) at the scale of the experiment, this will be driven towards non-perturbative regime through RGE at relatively low scales. Detailed discussion on this issue is out of the scope of this draft. 

\begin{figure}[thb]
$$
\includegraphics[scale=0.35]{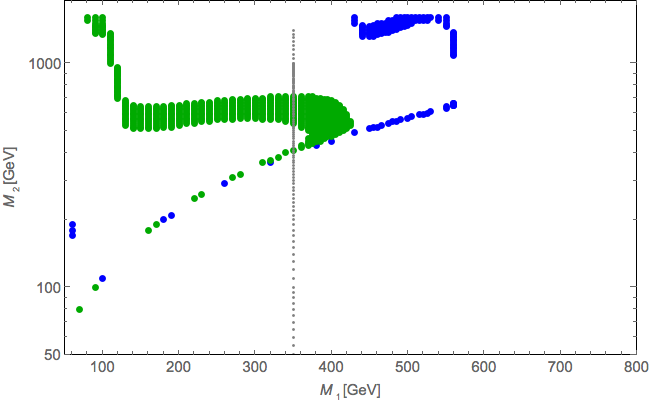}
\includegraphics[scale=0.35]{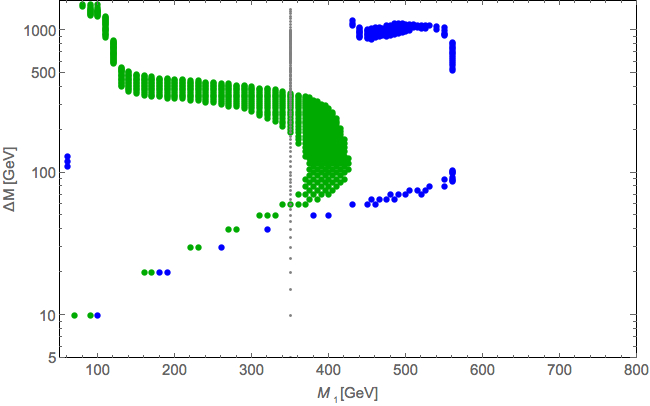}
$$
\caption{Left: $M_1$ versus $M_2$ (in GeV) scan for correct relic density. Right: $\Delta M$-$M_1$ scan for correct relic density. Both are obtained for  $\sin \theta=0.1,0.2$ 
in Blue and Green respectively. $M_Q=600$ GeV is chosen for illustration. The region to the right side of vertical dotted line denotes compatibility with 750 GeV excess.}
\label{fig:m1m2}
\end{figure}

The parameter space scan presented in this framework have been generated in the code MicrOmegas~\cite{Belanger:2008sj}, after implementing our DM model. 
In Fig. \ref{fig:Omegam-1}, we show how relic density changes with DM mass for different choices of mixing angle $\sin \theta=\{0.1,0.2, 0.3\}$, in blue, green and 
orange respectively with fixed mass difference $M_{2}-M_{1}=$ 100 GeV (left panel) and 500 GeV (right panel) with $M_Q=600$ GeV. It is clearly seen that the relic density 
drops significantly for DM mass larger than the vector like quark masses as the annihilation cross-section to those become very large with very large couplings through the 
scalar resonance $S$. To remind, we choose the couplings of both DM to $S$ and those of the vector like quarks to $S$ to be $5$ here for simplicity and from the requirement 
of diphoton excess. In Fig. \ref{fig:Omegam-2}, similar plots are shown for $M_Q=400$ GeV (left) and $M_Q>2000$ GeV (right) with $M_{2}-M_{1}=$ 100 GeV to show the 
dependence of the vector like quark masses on relic density. Thus we conclude that for DM mass $M_1 > M_Q$, we can not get the observed relic abundance while 
simultaneously explaining the diphoton excess which needs large coupling. On the other hand, $M_1 > M_S/2$ is required to prevent the tree-level decay of the resonance $S$. 
So for a given choice of $M_Q$, the allowed values of DM mass that can yield the correct relic density is $M_S/2 < M_1 < M_Q$. We then find the allowed parameter 
space within this mass range. We show the variation in relic density with DM mass for all possible mass splitting $\Delta M= M_2-M_1$ and for the same choices of the 
mixing angle $\sin \theta=\{0.1,0.2, 0.3\}$, in blue, green and orange respectively in Fig.  \ref{fig:Omegam-3} with $M_Q=600$ GeV. In all these plots, Figs (\ref{fig:Omegam-1}, 
\ref{fig:Omegam-2}, \ref{fig:Omegam-3}), the region to the right side of vertical dotted line denotes compatibility with 750 GeV excess. 
We see that for $\sin\theta\ge 0.3$, it is almost impossible to satisfy the relic density constraint. This is due to the fact that the large mixing leads to a larger cross-section 
of $\psi_1 \psi_1 \to W^+ W^- $ and always yields a low relic abundance. Therefore, in Fig. \ref{fig:m1m2}, we have shown the allowed parameter space for correct relic density 
using $\sin\theta=\{0.1, 0.2\}$ in terms of $\{M_1, M_2\}$ (left) and $\{M_1,\Delta M\}$ (right). With $\sin\theta=0.1$ (shown in Blue points in Fig. \ref{fig:m1m2}), due to small 
doublet fraction in the DM, the annihilations through $Z$ is smaller than required to satisfy relic density. Hence, coannihilation with $M_2$ is required to satisfy the relic abundance 
with smaller $\Delta M$ for $M_1$ upto 450 GeV. While for $\sin\theta =0.2$ (shown in Green points), annihilation cross-section itself (with larger $\Delta M$, which in turn enhance 
the Yukawa coupling) and annihilation plus coannihilation with smaller mass splitting ($\Delta M$), can yield required cross-section for correct density giving the allowed parameter 
space a funnel shape. Similar shape is obtained for $\sin \theta =0.1$ but with a larger DM mass.

\begin{figure}[thb]
$$\includegraphics[scale=0.40]{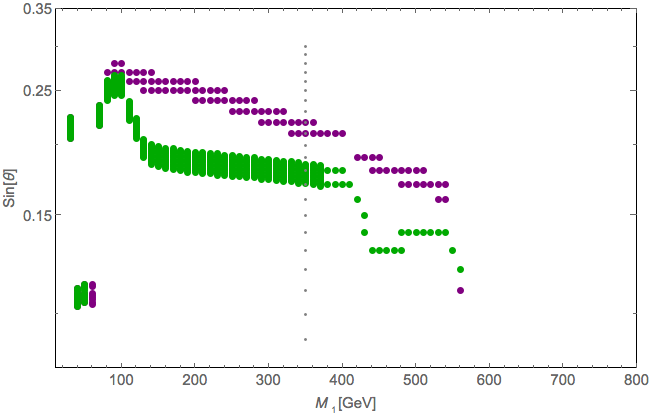}$$
\caption{ $\sin\theta$ versus $M_1$ scan for correct relic density. Purple: $\Delta M= 100$ GeV and Green: $\Delta M= 500$ GeV. $M_Q=600$ GeV is chosen 
for illustration purpose. The region to the right side of vertical dotted line denotes compatibility with 750 GeV diphoton excess.}
\label{fig:m1st}
\end{figure}
 
We have also demonstrated the  variation of DM mass with $\sin\theta$ for two fixed set of $\Delta M=\{100, 500\}$ GeV in Fig. \ref{fig:m1st}. 
We see that a moderately large region of the parameter space upto DM mass $M_1 < M_Q=600$ GeV can yield correct relic density. It is also worthy to mention that there are other possible annihilations of the DM, for example, to $hS, SS$ etc, which doesn't affect the phenomenology of obtaining correct relic density to a great extent as we have chosen  $M_Q<M_S$ and the annihilation to vector-like quark pair dominates over the others whenever they are kinematically allowed.

\begin{figure}[thb]
$$\includegraphics[scale=0.35]{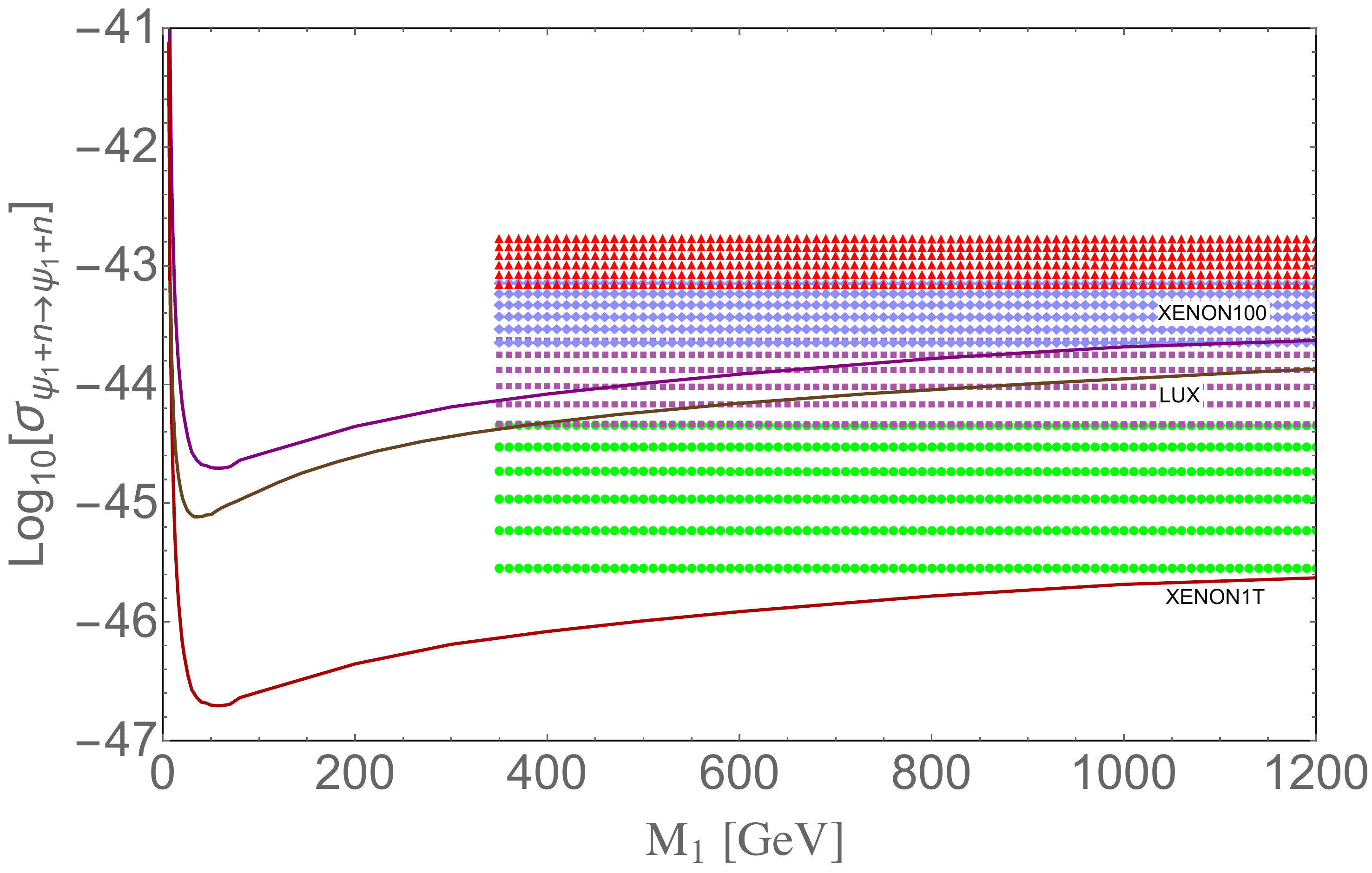}$$
\caption{ Spin independent direct detection cross-section for $\psi_1$ DM as a function of its mass. Constraints from XENON 100, LUX data and predictions 
at XENON 1T for the DM are shown in thick lines. Green ($\sin\theta = 0.05-0.1$), purple ($\sin\theta = 0.1 - 0.15$), lilac ($\sin\theta = 0.15 - 0.2$) and red 
($\sin\theta = 0.2 - 0.25$) regions are shown respectively from bottom to top.}
\label{fig:DD2}
\end{figure}

Direct detection of this DM occurs through  $Z$ and $h$ mediation. We show the spin-independent cross-section for $\psi_1$ DM by taking its mass 
range $M_{1}: 375-1200$ GeV and varying $\sin\theta=\{0.05-0.25\}$ in Fig. \ref{fig:DD2}. Green ($\sin\theta = 0.05-0.1$), purple ($\sin\theta = 0.1 - 0.15$), lilac ($\sin\theta 
= 0.15 - 0.2$) and red ($\sin\theta = 0.2 - 0.25$) regions are shown respectively from bottom to top. It clearly shows that lower values of $\sin\theta$ is allowed 
while the larger ones are discarded. Together, $\sin\theta \le 0.15$ is allowed by LUX data which also coincides with correct relic density search.

\section{Conclusions}\label{conclusions}
We have illustrated how the recent diphoton excess signal $pp \to S \to \gamma \gamma$ around an invariant mass of $750~$GeV can be accounted by a Dark sector 
assisted scalar decay. The framework considered is a simple extension of SM with additional scalar singlet $S$ having mass around $750~$GeV, an iso-singlet 
vector-like quark $Q$ and a dark sector constituted by a vector-like lepton doublet $\psi$ and a neutral singlet $\chi^0$. We argue that the extra particles added in this framework 
are minimal when we explain diphoton excess signal and DM component of the Universe. We note that the masses of the new particles added are below TeV scale, 
but above $M_S/2=375$ GeV. We found that correct relic density can be obtained for DM mass ($M_1$) within the range: $M_S/2 < M_1 < M_Q$ and the mixing 
in the dark sector require to be small $\sin \theta \le 0.1$ from direct search constraints. Since $\psi$ is heavy, it evades constraints coming from oblique corrections 
~\cite{Arina:2012aj}. The DM remains elusive at collider. However, its charged partner $\psi^\pm$ can be searched at LHC. In particular, if the 
singlet-doublet mixing is very small ($\sim 10^{-5}$) then the charged partner of the DM, which assist for diphoton excess, can give rise a displaced 
vertex signature \cite{Bhattacharya:2015qpa}.

\section{Acknowledgements}
The work of SB is partially supported by DST INSPIRE grant no PHY/P/SUB/01 at IIT Guwahati. 
The work of SP is partly supported by DST, India under the financial grant SB/S2/HEP-011/2013. 
Narendra Sahu is partially supported by the Department of Science and Technology, Govt. of India under 
the financial Grant SR/FTP/PS-209/2011. 


\bibliographystyle{JHEP}
\end{document}